\documentclass[american,aps,pra,reprint,superscriptaddress,nofootinbib]{revtex4-2}
\usepackage[T1]{fontenc}
\usepackage[latin9]{inputenc}
\usepackage{color}
\usepackage{graphicx}
\usepackage{babel}
\usepackage{amsmath}
\usepackage{amsthm}
\usepackage{amssymb}
\usepackage{physics}
\usepackage{appendix}
\usepackage[unicode=true,pdfusetitle,
 bookmarks=true,bookmarksnumbered=false,bookmarksopen=false,
 breaklinks=false,pdfborder={0 0 0},pdfborderstyle={},backref=false,colorlinks=true]
 {hyperref}
\hypersetup{
 allcolors=magenta}

\makeatletter
\theoremstyle{plain}
\newtheorem{thm}{\protect\theoremname}
\ifx\proof\undefined
\newenvironment{proof}[1][\protect\proofname]{\par
	\normalfont\topsep6\p@\@plus6\p@\relax
	\trivlist
	\itemindent\parindent
	\item[\hskip\labelsep\scshape #1]\ignorespaces
}{%
	\endtrivlist\@endpefalse
}
\providecommand{\proofname}{Proof}
\fi
\newtheorem{proposition}{Proposition}

\usepackage{times}
\usepackage{txfonts}
\usepackage{braket}
\usepackage{colortbl}

\makeatother

\providecommand{\theoremname}{Theorem}

\begin{document}
\title{Local Purity Distillation in Quantum Systems: \\ Exploring the Complementarity Between Purity and Entanglement}

\author{Ray Ganardi}
\affiliation{Centre for Quantum Optical Technologies, Centre of New Technologies,
University of Warsaw, Banacha 2c, 02-097 Warsaw, Poland}

\affiliation{School of Physical and Mathematical Sciences, Nanyang Technological University, 21 Nanyang Link, Singapore, 637371
}
\author{Piotr Masajada}
\affiliation{Institute of Fundamental Technological Research, Polish Academy of Sciences, \\ Pawi\'nskiego 5B, 02-106 Warsaw, Poland}

\author{Moein Naseri}
\affiliation{Centre for Quantum Optical Technologies, Centre of New Technologies,
University of Warsaw, Banacha 2c, 02-097 Warsaw, Poland}

\author{Alexander Streltsov}
\email{streltsov.physics@gmail.com}
\affiliation{Institute of Fundamental Technological Research, Polish Academy of Sciences, \\ Pawi\'nskiego 5B, 02-106 Warsaw, Poland}

\begin{abstract}
    Quantum thermodynamics and quantum entanglement represent two pivotal quantum resource theories with significant relevance in quantum information science. Despite their importance, the intricate relationship between these two theories is still not fully understood. Here, we investigate the interplay between entanglement and thermodynamics, particularly in the context of local cooling processes. We introduce and develop the framework of Gibbs-preserving local operations and classical communication. Within this framework, we explore strategies enabling remote parties to effectively cool their local systems to the ground state. Our analysis is centered on scenarios where only a single copy of a quantum state is accessible, with the ideal performance defined by the highest possible fidelity to the ground state achievable under these constraints. We focus on systems with fully degenerate local Hamiltonians, where local cooling aligns with the extraction of local purity. In this context, we establish a powerful link between the efficiency of local purity extraction and the degree of entanglement present in the system, a concept we define as \emph{purity-entanglement complementarity}. Moreover, we demonstrate that in many pertinent scenarios, the optimal performance can be precisely determined through semidefinite programming techniques. Our findings open doors to various practical applications, including techniques for entanglement detection and estimation. We demonstrate this by evaluating the amount of entanglement for a class of bound entangled states. 
\end{abstract}

\maketitle

\section{Introduction}
Quantum resource theories serve as a
fundamental framework for probing quantum phenomena and steering advancements
in quantum technologies~\cite{RevModPhys.91.025001}. Foremost among these theories stand the studies
of quantum entanglement~\cite{RevModPhys.81.865} and quantum thermodynamics~\cite{PhysRevLett.111.250404,Horodecki2013}. The resource theory of entanglement investigates the utility of entangled quantum systems within local
constraints. Conversely, the resource theory of thermodynamics assesses
the potential and limitations of manipulating quantum systems when bound by energy constraints.

In any quantum resource theory, a pivotal challenge is the quantification
of resources, which entails gauging the resource content in a specific
quantum state. Such resource quantifiers play an integral role in determining
a quantum state's aptitude for quantum information processing tasks.
In the realm of entanglement quantification, it is essential that
the quantifier does not increase under local operations and classical communication
(LOCC)~\cite{PhysRevLett.78.2275}. In the bipartite context, a plethora of entanglement quantifiers
have been proposed, most of which can be computed efficiently for
pure bipartite states~\citep{RevModPhys.81.865}. However, when it
comes to multipartite scenarios and noisy states, the task becomes substantially more
complex, with numerous established entanglement quantifiers proving
challenging to compute efficiently, even for pure states. 

Concurrently, another salient issue in quantum resource theories is
discerning the most efficient methods to create valuable resource
states. This concern emerges from the inevitability of noise in practical
implementations. Although the crux of quantum information processing
hinges on pure quantum states, the quintessential quantum state encountered
in a laboratory setting is invariably noisy, stemming from unavoidable
environmental interactions. Hence, it is essential to develop methodologies to convert noisy states into pure resourceful states.

In this article, we introduce the concept of \emph{Gibbs-preserving
local operations and classical communication (GLOCC)}. Specifically,
these are LOCC protocols that preserve the thermal state of a bipartite or
multipartite system. Within this framework, we investigate optimal procedures to cool local systems, corresponding to bringing them to their local ground state. When the local Hamiltonians are fully degenerate, Gibbs-preserving operations give rise to the resource theory of purity~\cite{PhysRevA.67.062104,GOUR20151,Streltsov_2018}. Under these conditions, the quest for local cooling
mirrors the endeavor to distill pure states at the local level. We establish an intricate relationship between the locally extractable purity and entanglement amount, a phenomenon we coin as \emph{purity-entanglement
complementarity}. As a quantifier of entanglement we use geometric entanglement~\cite{PhysRevA.68.042307,Streltsov_2010}, which is useful both for bipartite and multipartite entanglement quantification. 
\begin{figure*}
\includegraphics[width=0.7\textwidth]{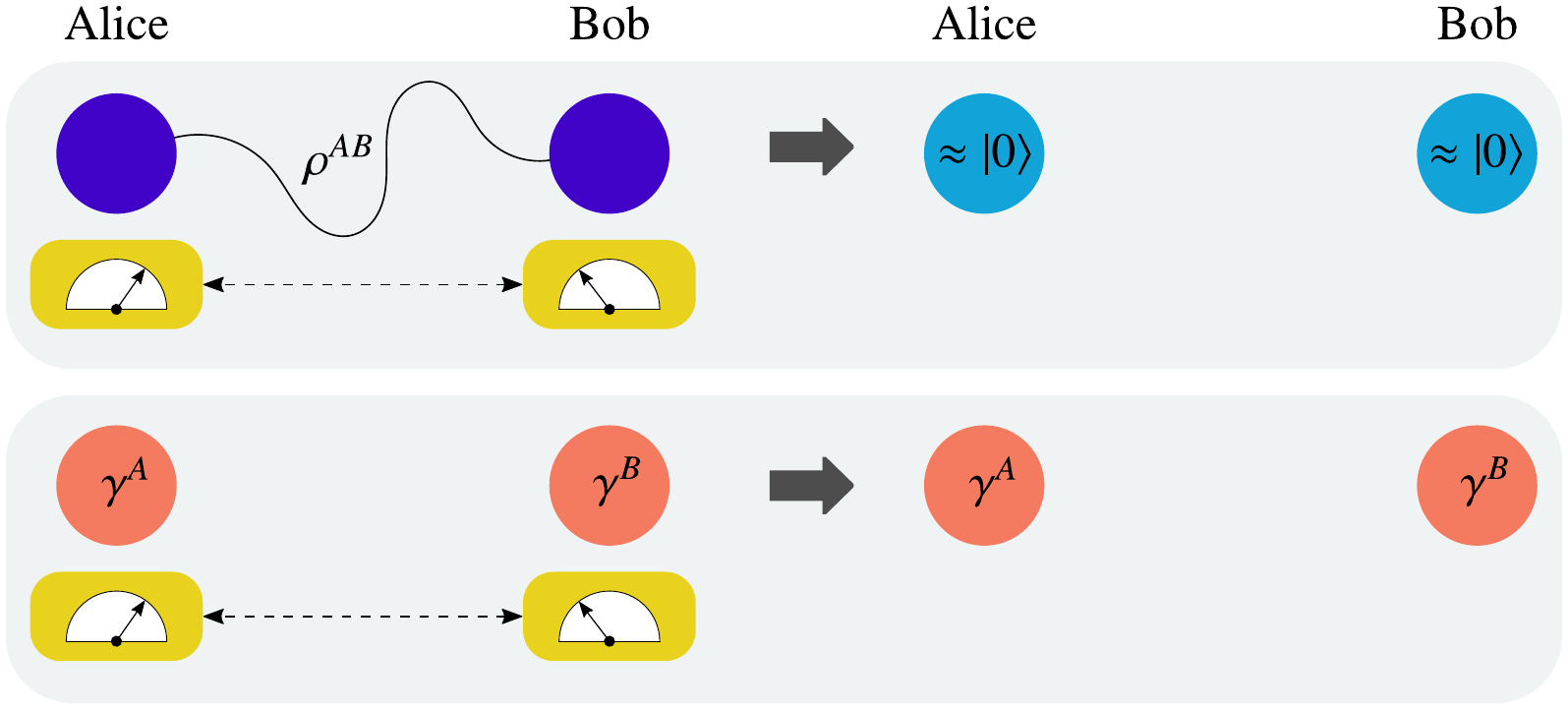}

\caption{\label{fig:LocalCooling}Local cooling of a bipartite system by local Gibbs-preserving operations and classical communication. The goal of the process is to convert an initial bipartite state $\rho^{AB}$ into the ground state $\ket{00}$ with the maximal fidelity (upper part of the figure). The procedure preserves the total Gibbs state $\gamma^A\otimes \gamma^B$ (lower part of the figure).}

\end{figure*}

Utilizing semidefinite programming (SDP) techniques, we formulate approaches
to estimate the optimal fidelity for establishing local ground states. The delineated relationship between purity and entanglement further allows our techniques to detect and quantify entanglement in numerous relevant settings, if the entanglement is quantified via geometric entanglement. In the bipartite setting, we can estimate the geometric entanglement of any rank-2 state if one of the subsystems is a qutrit, and for any rank-3 state if one of the subsystems is a qubit. Our techniques can also be used to estimate geometric entanglement of PPT entangled states and certain families of multipartite pure states.

\section{Local cooling with Gibbs-preserving LOCC}

In this work, we investigate a context where access is restricted
to a single copy of a quantum state. We will first briefly discuss the setting for a single party. Given a quantum state $\rho$, our objective is to cool the system to the ground state $\ket{0}$ using Gibbs-preserving
operations~\cite{Faist_2015}. These are transformations which preserve
the Gibbs state $\gamma=e^{-\beta H}/\mathrm{Tr}[e^{-\beta H}]$,
where $H$ is the Hamiltonian of the system and $\beta=\frac{1}{kT}$ signifies
the inverse temperature. Perfectly attaining the ground state is generally unfeasible; thus, our interest centers on maximizing the fidelity with the ground state. This optimal fidelity is defined as $\max_{\Lambda}\braket{0|\Lambda[\rho]|0}$, and the maximum is taken over all Gibbs-preserving operations $\Lambda$. In the presence of a fully degenerate Hamiltonian, the maximal fidelity coincides
with the largest eigenvalue of $\rho$, denoted as $\lambda_{\max}(\rho)$.

Transitioning to a bipartite scenario involving two parties, Alice
and Bob, each has local Hamiltonians $H_{A}$ and $H_{B}$ respectively.
The collective Hamiltonian is represented as $H=H_{A}+H_{B}$. The
associated Gibbs state is given by $\gamma=\gamma^{A}\otimes\gamma^{B}$,
where $\gamma^{X}$ denotes the local Gibbs state of the party $X$.
Drawing an analogy to Gibbs-preserving operations, an LOCC protocol $\Lambda$ is termed Gibbs-preserving if $\Lambda[\gamma]=\gamma$. The set of all such transformations will be denoted by GLOCC.

With local energy conceptualized as a resource, two distinct scenarios
emerge. In the first scenario, Alice and Bob jointly possess a state
$\rho^{AB}$ and their objective is to jointly cool their systems to the ground state $\ket{00}$, see also Fig.~\ref{fig:LocalCooling}. The figure of merit in this scenario
is given by 
\begin{equation}
F_{\max}(\rho^{AB})=\max_{\Lambda\in\text{GLOCC}}\braket{00|\Lambda[\rho^{AB}]|00}. \label{eq:FmaxDefinition}
\end{equation}
In the second scenario, Alice teams up with Bob to cool his individual
system to the ground state. The figure of merit is then given
by
\begin{equation}
F^{A|B}(\rho^{AB})=\max_{\Lambda\in\text{GLOCC}}\braket{0|\mathrm{Tr}_{A}(\Lambda[\rho^{AB}])|0}.
\end{equation}
For the second scenario, we will also consider the setting where the
classical communication is directed from Alice to Bob only. The corresponding
figure of merit will be denoted by $F_{\rightarrow}^{A|B}$. 

A noteworthy situation arises when the local Hamiltonians associated
with Alice and Bob are fully degenerate. Here, the Gibbs-preservation
requirement equates to the LOCC protocol preserving the maximally
mixed state. The task of establishing a ground state is then equivalent
to extraction of local purity in this setting. 

In this article, our emphasis is primarily on the single-copy scenario. However, it is noteworthy that related settings within the asymptotic regime have been explored in previous works, as detailed in references~\cite{PhysRevLett.89.180402,PhysRevA.71.062307,PhysRevLett.122.130601}. Single-shot purity distillation in the distributed setting has also been considered recently in \cite{chakraborty2022oneshot}. For thermal operations, which is a subclass of Gibbs-preserving operations, a related setting has been considered in~\cite{PhysRevA.95.012313}.

\section{Optimal extraction of local purity}
In the subsequent discussion, we investigate techniques to evaluate
$F_{\max}$, focusing on the setting where the local Hamiltonians are fully degenerate. Under this premise, $F_{\max}$ can be succinctly represented. 
\begin{proposition}\label{prop:local-purity}
  When the local Hamiltonians are fully degenerate, we have
\begin{equation}
F_{\max}(\rho^{AB})=\max_{U_{A},V_{B}}\braket{00|U_{A}\otimes V_{B}\rho^{AB}U_{A}^{\dagger}\otimes V_{B}^{\dagger}|00}. \label{eq:Fmax}
\end{equation}
\end{proposition}

Hence, to optimize fidelity with the ground state, the best strategy
involves implementing local unitaries for both Alice and Bob. This
finding possesses relevance even when extended to setups with more
than two parties. For a detailed exposition on this, the reader is
directed to the Appendix.

A straightforward
examination yields the following upper bound: 
\begin{equation}
F_{\max}(\rho^{AB})=\max_{\sigma^{AB}\in\mathrm{SEP}}\mathrm{Tr}(\rho^{AB}\sigma^{AB})\leq\max_{\sigma^{AB}\in\mathrm{PPT}}\mathrm{Tr}(\rho^{AB}\sigma^{AB}), \label{eq:FmaxPPT}
\end{equation}
where SEP and PPT denote separable states, and states
that exhibit positive partial transpose, respectively. Note that the maximization over PPT states is feasible via SDP. If the dimensions of $A$
and $B$ are such that $d_{A}d_{B}\leq6$, the inequality~(\ref{eq:FmaxPPT})
becomes equality, due to the fact that the sets SEP and PPT coincide in these cases~\cite{HORODECKI19961}. This allows to evaluate $F_{\max}$ in all such settings.
In general, when the state $\sigma^{*} \in \mathrm{PPT}$ that maximizes Eq.~(\ref{eq:FmaxPPT})
is pure, the bound in Eq.~(\ref{eq:FmaxPPT}) is tight. 

By performing the maximization over PPT states outlined in Eq.~(\ref{eq:FmaxPPT}) and discerning
the largest eigenvalue of the optimal $\sigma^{*} \in \mathrm{PPT}$, one can approximate
$F_{\max}$ with high precision in various settings. Elaborating further, for $\lambda_{\max}(\sigma^{*})=1-\varepsilon$,
we establish the following proposition.

\begin{proposition}\label{prop:optimizer-purity}
  It holds that
  \begin{equation}
    \mathrm{Tr}(\rho\sigma^{*})+\delta\geq F_{\max}(\rho)\geq\mathrm{Tr}(\rho\sigma^{*})-4\sqrt{\varepsilon}, \label{eq:FmaxSDP}
  \end{equation}
  where $\delta$ is the precision to which the SDP in Eq. (\ref{eq:FmaxPPT}) is computed numerically.
\end{proposition}

Empirically, we found that $\varepsilon$ can often be made close to zero, which allows to evaluate $F_{\max}$ with an arbitrary
precision in practice. In more detail, for bipartite system of dimension $d_A=d_B=3$, we sampled $2 \times 10^{5}$ random mixed states\footnote{To sample a bipartite mixed state of dimension $d_{AB}$, we sample Haar random pure states of dimension $d_{AB}^{2},$ and trace out a subsystem of dimension $d_{AB}.$}, finding that in 13 cases $\varepsilon$ was above $0.08$, and in all other cases $\varepsilon$ was below $2 \times 10^{-9}$. For systems of dimension $d_A=d_B=4$ we sampled $2 \times 10^{5}$ random mixed states, finding that in 31 cases $\varepsilon$ was above $8\times10^{-4},$ and in all other states $\varepsilon$ was below $3\times10^{-8}.$ For $d_A=d_B=5$ we sampled $3\times10^{4}$ random mixed states, finding that for 23 of them $\varepsilon$ was above $0.005,$ and for all other states $\varepsilon$ was below $5\times10^{-9}.$
These results also extend
to the multipartite setting, we refer to the Appendix for more details, as well as the proof of the proposition.

Let us now explore the second scenario where Alice aids Bob in cooling
his system to its ground state. We initially consider a framework
where classical communication flows exclusively from Alice to Bob.

When Alice executes a local measurement characterized by POVM elements
$\{M_{i}^{A}\}$, the probability of obtaining measurement outcome
$i$ manifests as: $p_{i}=\mathrm{Tr}[M_{i}^{A}\otimes\openone^{B}\rho^{AB}]$.
Given Alice's measurement outcome, the resultant state of Bob's system
is: $\sigma_{i}^{B}=\mathrm{Tr}_{A}[M_{i}^{A}\otimes\openone^{B}\rho^{AB}]/p_{i}$.
The maximal fidelity Bob can achieve with the ground state in this
setup is represented as: 
\begin{equation}
F_{\rightarrow}^{A|B}(\rho^{AB})=\max\sum_{i}p_{i}\lambda_{\max}(\sigma_{i}^{B}),
\end{equation}
with the optimization extending over all ensembles $\{p_{i},\sigma_{i}^{B}\}$
achievable from $\rho^{AB}$ by Alice's measurements, and $\lambda_{\max}(\sigma)$ denoting the largest eigenvalue of $\sigma$. Collating these
results, $F_{\rightarrow}^{A|B}$ is expressed as: 
\begin{equation}
F_{\rightarrow}^{A|B}(\rho^{AB})=\max\sum_{i}\mathrm{Tr}\left[M_{i}^{A}\otimes\mu_{i}^{B}\rho^{AB}\right],\label{eq:Pa-1}
\end{equation}
where we optimize over all POVMs $\{M_{i}^{A}\}$ and quantum states
$\{\mu_{i}^{B}\}$.

While in the above analysis we assumed that Alice can perform arbitrary POVMs locally, note that Alice can implement a general POVM $\{M_{i}^{A}\}$ via a unital operation having the Kraus operators
\begin{equation}
K_{i}^{A}=\sqrt{M_{i}^{A}}.
\end{equation}
Thus, optimal performance as in Eq.~(\ref{eq:Pa-1}) can also be achieved via GLOCC with one-way communication.

To provide an upper bound for $F_{\rightarrow}^{A|B}$, we can state:
\begin{equation}
F_{\rightarrow}^{A|B}(\rho^{AB})\leq\max_{X^{AB}}\mathrm{Tr}\left[X^{AB}\rho^{AB}\right],\label{eq:AssistedSDP-1}
\end{equation}
where the optimization is over all matrices $X^{AB}$ satisfying: 
\begin{equation}
X^{AB}\geq0,\quad X^{T_{B}}\geq0,\quad X^{A} = \openone_{A}.\label{eq:AssistedSDP-2}
\end{equation}
It is important to note that the right-hand side of Eq.~(\ref{eq:AssistedSDP-1})
is computationally feasible via SDP. As we will see in the following,
for systems of small dimension the inequality~(\ref{eq:AssistedSDP-1})
is tight. 
\begin{thm}
\label{thm:F}For $d_{A}d_{B}\leq6$ it holds 
\begin{equation}
F_{\rightarrow}^{A|B}(\rho^{AB})=\max_{X^{AB}}\mathrm{Tr}\left[X^{AB}\rho^{AB}\right]\label{eq:AssistedSDP-3}
\end{equation}
with $X^{AB}$ fulfilling Eqs.~(\ref{eq:AssistedSDP-2}). 
\end{thm}
\noindent We refer to the Appendix for the proof.

We will now turn our attention to the more general setting where classical communication is possible in both directions. For $d=d_{A}=d_{B}$ we
will show that the problem is symmetric under permutation of Alice
and Bob. If Alice assists Bob to extract purity, an optimal GLOCC
protocol can always be assumed to produce a state of the
form $\openone/d\otimes\sigma^{B}$, where the state $\sigma^{B}$ is diagonal in the
computational basis. We will now show that by using GLOCC operations
it is possible to convert this state into $\sigma^{A}\otimes\openone/d$. Denoting the eigenvalues
of $\sigma$ with $p_{i}$,
we have $\openone/d\otimes\sigma=1/d\sum_{ij}p_{i}\ket{ji}\!\bra{ji}$
and $\sigma\otimes\openone/d=1/d\sum_{ij}p_{i}\ket{ij}\!\bra{ij}$.
Alice and Bob can now apply the following GLOCC protocol to convert
$\openone/d\otimes\sigma$ into $\sigma\otimes\openone/d$. For this, Alice
and Bob perform local measurements in the computational basis and
send the outcome of the measurement to the other party. If Alice obtains
outcome $j$, and Bob obtains outcome $i$, then Alice applies a local
unitary transforming $\ket{j}$ into $\ket{i}$. Correspondingly,
Bob applies a local unitary which transforms $\ket{i}$ into $\ket{j}$.
It is straightforward to check that the overall transformation is
GLOCC, since it preserves the total maximally mixed state. These arguments also demonstrate that two-sided communication is in general useful in the assisted setting, leading to a better performance of the local cooling procedure.

We will now present a method for approximating the optimal fidelity
$F^{A|B}(\rho^{AB})$, if classical communication is allowed in both directions. In a very similar way as in Eq.~(\ref{eq:AssistedSDP-1})
we can obtain an upper bound for $F^{A|B}(\rho^{AB})$.

\begin{proposition}~\label{prop:assisted-sdp}
It holds that
  \begin{equation}
  F^{A|B}(\rho^{AB})\leq\max_{X^{AB}}\mathrm{Tr}\left[X^{AB}\rho^{AB}\right],\label{eq:AssistedSDP-5}
\end{equation}
where the optimization is over all $X^{AB}$ satisfying
\begin{equation}
\openone\geq X^{AB}\geq0,\,\,\,\,\,X^{T_{B}}\geq0,\,\,\,\,\,\mathrm{Tr}[X^{AB}]=d_A. \label{eq:AssistedSDP-6}
\end{equation}
\end{proposition}

\noindent The proof of this statement is given in the Appendix.

\section{Geometric entanglement and purity-entanglement complementarity} 

In the following, we will review the main properties of geometric entanglement, an entanglement quantifier which is applicable in both, bipartite and multipartite settings~\cite{PhysRevA.68.042307,Streltsov_2010}. 
For a pure $n$-partite state $\psi=\ket{\psi}\!\bra{\psi}$, the
multipartite geometric entanglement is defined as~\cite{Shimony1995,PhysRevA.68.042307,HBarnum2001,PhysRevA.65.062312}
\begin{equation}
E_{g}(\psi)=1-\max_{\phi\in\mathrm{SEP}}\left|\braket{\phi|\psi}\right|^{2},
\end{equation}
where SEP denotes the set of fully separable states. For
mixed states, the geometric entanglement is defined as~\cite{PhysRevA.68.042307} 
\begin{equation}
E_{g}(\rho)=\min\sum_{i}p_{i}E_{g}(\psi_{i}),
\end{equation}
where the minimum is taken over all pure state decompositions of $\rho$
such that $\sum_{i}p_{i}\psi_{i}=\rho$. As has been shown in~\cite{Streltsov_2010},
$E_{g}$ can also be expressed as 
\begin{equation}
E_{g}(\rho)=1-\max_{\sigma\in\mathrm{SEP}}F(\rho,\sigma)
\end{equation}
with fidelity $F(\rho,\sigma)=(\mathrm{Tr}\sqrt{\sqrt{\rho}\sigma\sqrt{\rho}})^{2}$. The geometric entanglement is zero for all separable states, and positive whenever the state is entangled. Moreover, $E_g$ does not increase under LOCC~\cite{PhysRevA.68.042307}. Several related entanglement quantifiers for mixed states have also been proposed in the literature~\cite{PhysRevA.57.1619,PhysRevA.73.044301}.

We will now demonstrate a powerful complementarity relations, linking
the amount of geometric entanglement of a quantum state to the maximal
fidelity with the ground state, achievable in an assisted scenario with one-way communication. 
\begin{thm}
\label{thm:Complementarity}For any pure state $\ket{\psi}^{ABC}$
it holds that 
\begin{equation}
F_{\rightarrow}^{A|B}(\rho^{AB})+E_{g}(\rho^{BC})=1.
\end{equation}
\end{thm}
\noindent The proof of the theorem can be found in the Appendix.

The results presented above provide a strong link between the task of assisted purity extraction and the geometric entanglement, and can also be seen as an operational interpretation of the geometric entanglement in a thermodynamical setting. Furthermore, this relation also holds for infinite dimensional systems.

A similar complementarity relation can also be found for the maximal
local purity $F_{\max}(\rho^{AB})$ of a bipartite state and the tripartite
geometric entanglement of the purification $\ket{\psi}^{ABC}$. In
particular, it holds that $E_{g}(\psi^{ABC})+F_{\max}(\rho^{AB})=1$, as follows directly from results in~\cite{PhysRevA.77.062317}.
This result extends to any number of parties. For an $n$-partite
state $\rho_{n}$ let $\ket{\psi_{n+1}}$ be its $n+1$-partite purification.
Then, it holds that~\cite{PhysRevA.77.062317} $E_{g}(\psi_{n+1})+F_{\max}(\rho_{n})=1$.

\section{Applications} 

The methods presented in this work
have various applications, which we will discuss in the following.
One such application concerns evaluation of geometric entanglement.
Due to the complementarity relation in Theorem~\ref{thm:Complementarity},
evaluation of geometric entanglement of a mixed state is equivalent
to the evaluation of $F_{\rightarrow}$ on a purifying system. Recall
that we can evaluate $F_{\rightarrow}^{A|B}$ for any system with
$d_{A}d_{B}\leq6$ via SDP, see Theorem~\ref{thm:F}. This means
that the geometric entanglement of any state $\rho^{AB}$ can be evaluated
via SDP, whenever one of the subsystems is a qutrit and the state
has rank at most 2. If one of the subsystems is a qubit, the evaluation
of geometric entanglement is possible for all states with rank at
most 3. Interestingly, there is no limit on the dimension of the second
subsystem. To our knowledge, the geometric entanglement is the first
faithful entanglement measure which can be evaluated via SDP for all
such states. When it comes to evaluation of geometric entanglement
of pure states, our methods imply that $E_{g}(\ket{\psi}^{ABC})$
can be evaluated via SDP whenever $d_{A}d_{B}\leq6$, i.e., when the
joint dimension of any two subsystems is at most $6$. 

\begin{figure}
\begin{centering}
\includegraphics[width=\columnwidth]{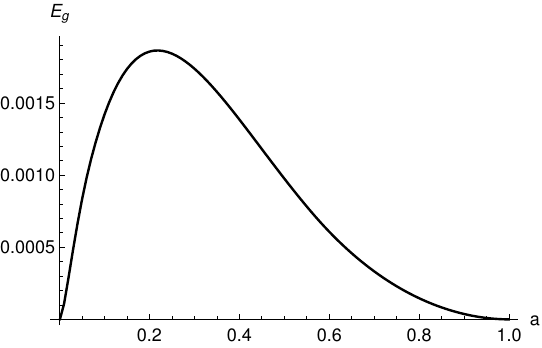}
\par\end{centering}
\caption{\label{fig:BE} Geometric entanglement for the $2 \times 4$ PPT entangled states $\rho_a$, see Supplemental Material for the definition.}
\end{figure}

In more general cases, our methods provide a lower bound on the geometric entanglement which is feasible via SDP. This is a consequence of Theorem~\ref{thm:Complementarity}
and the SDP upper bound on $F_{\rightarrow}$ provided in Eq.~(\ref{eq:AssistedSDP-1}). In fact, this approach also leads to the following lower bound on the geometric entanglement: 
\begin{equation}
E_{g}(\rho^{BC})\geq1-\max_{X^{ABC}}\mathrm{Tr}\left[X^{ABC}\psi^{ABC}\right],\label{eq:EgLowerBound-2}
\end{equation}
where $\ket{\psi}^{ABC}$ is a purification of $\rho^{BC}$ and the maximum is taken over all $X^{ABC}$ with the properties
\begin{equation}
X^{ABC}\geq0,\,\,\,X^{T_{A}}\geq0,\,\,\,X^{T_{B}}\geq0,\,\,\,X^{T_{C}}\geq0,\,\,\,X^{A}=\openone_{A}.
\end{equation}
We refer to the Appendix for more details. This lower bound is nonzero even for some bound entangled
state, as we demonstrate in Fig.~\ref{fig:BE} for the $2 \times 4$ bound entangled Horodecki states. Surprisingly, our numerical results indicate that the lower bound is tight in this case. In more detail, we find numerically that this lower bound matches with the upper bound on geometric entanglement which we obtained using the algorithm in~\cite{PhysRevA.84.022323}. The numerical gap between the upper and the lower bound is $\Delta E_g \leq 4 \times 10^{-7}$, and the average gap is $\overline{\Delta E_{g}}\approx1.2\times10^{-8}$, further details can be found in the Appendix. Our results complement earlier research on approximating separable states and evaluating entanglement of bipartite and multipartite states~\cite{Eisert_2004,10.1145/1993636.1993683,10.1007/978-3-642-31594-7_67,Wei_2010,Orus_2010,Zhang_2020,Harrow_2017,Hua_2016,Doherty_2005,Teng_2017,PhysRevA.100.062318,PhysRevApplied.13.054022}.

\section{Conclusions}
We have explored the intersection of quantum thermodynamics and quantum entanglement, advancing our understanding of these quantum resource theories. Our development of the Gibbs-preserving local operations and classical communication (GLOCC) framework has enabled us to investigate the intricacies of local cooling processes in quantum systems. This framework has proven to be a powerful tool in analyzing the efficiency and strategies for cooling local systems to their ground states, particularly in settings where only a single copy of the quantum state is available.

In scenarios where local Hamiltonians are fully degenerate, the process of cooling a local system to its ground state corresponds to the extraction of local purity. For this setting, we have establishes the concept of purity-entanglement complementarity, illustrating a profound connection between the extraction of local purity and the amount of entanglement within a system. This relationship not only offers a practical and operational interpretation of geometric entanglement within the realm of thermodynamics but also simplifies and enhances the evaluation of geometric entanglement.

Our research has shown that the optimal performance of the cooling process, specifically the maximum fidelity achievable with local ground states, can be efficiently determined through semidefinite programming across a wide range of pertinent scenarios. Leveraging the concept of purity-entanglement complementarity, we have further extended this capability to accurately estimate the geometric entanglement in various families of multipartite pure states as well as bipartite mixed states. 

The methodologies introduced in this study hold the potential to tackle a range of pertinent questions in quantum information science, particularly those which can be linked to the evaluation of geometric entanglement. Important examples are verifying the positivity of linear maps and evaluating the maximal output purity of quantum channels~\cite{Zhu_2010}. While a comprehensive analysis of these applications is beyond the scope of this article, they present compelling directions for future investigations.

Another compelling question which is left open in this article is related to the effectiveness of the entanglement-detection and estimation method emerging from the purity-entanglement complementarity. While we have successfully shown that this approach can estimate entanglement in specific PPT entangled states with high precision, the performance of this method for other sets of states is currently unclear. Drawing on the findings documented in this work, we hypothesize that the methods developed here could be successfully applied to more general families of noisy states not explicitly addressed in this article.

\acknowledgements
We thank Bartosz Regula and Varun Narasimhachar for discussion. This work was supported by the ``Quantum Optical Technologies'' project, carried out within the International Research Agendas programme of the Foundation for Polish Science co-financed by the European Union under the European Regional Development Fund and the National Science Centre Poland (Grant No. 2022/46/E/ST2/00115) and within the QuantERA II Programme (Grant No. 2021/03/Y/ST2/00178, acronym ExTRaQT) that has received funding from the European Union's Horizon 2020 research and innovation programme under Grant Agreement No. 101017733.

\bibliography{literature}

\appendix

\section{Proof of Proposition~\ref{prop:local-purity}}
We will show that $F_{\max}$ as defined in Eq.~(\ref{eq:FmaxDefinition}) of the main text can also be expressed as in Eq.~(\ref{eq:Fmax}) of the main text. For this, recall that any LOCC protocol $\Lambda$ also corresponds to a separable operation $\Lambda(\rho)=\sum_{i}K_{i}\rho K_{i}^{\dagger}$ with Kraus operators of the form $K_i = A_i \otimes B_i$~\cite{PhysRevA.57.1619}. If the operation additionally preserves the maximally mixed state, then $\{K_{i}^{\dagger}\}$ also correspond to a valid set of Kraus operators, giving rise to the transformation $\Lambda^\dagger(\rho)=\sum_{i}K_{i}^{\dagger}\rho K_{i}$. In this case, $\Lambda^\dagger$ is also a separable operation which preserves the maximally mixed state. Denoting with GSEP the set of separable operations which preserve the maximally mixed state we obtain the following:
\begin{align}
F_{\max}(\rho^{AB}) & \leq\max_{\Lambda\in\text{GSEP}}\braket{00|\Lambda[\rho^{AB}]|00}\\
 & =\max_{\Lambda^\dagger\in\text{GSEP}}\mathrm{Tr}\left[\rho^{AB}\Lambda^\dagger\left(\ket{00}\!\bra{00}\right)\right]\nonumber \\
 & \leq\max_{\ket{\psi},\ket{\phi}}\mathrm{Tr}\left[\rho^{AB}\ket{\psi}\!\bra{\psi}^{A}\otimes\ket{\phi}\!\bra{\phi}^{B}\right].\nonumber 
\end{align}
The last inequality follows from the fact that $\Lambda^\dagger(\ket{00}\!\bra{00})$ is a separable state. It is further clear that equality can be achieved by choosing $\Lambda^\dagger$ to be a product of two local unitaries, i.e., $\Lambda^\dagger(\rho)=U_{A}\otimes V_{A}\rho^{AB}U_{A}^{\dagger}\otimes V_{A}^{\dagger}$. This completes the proof of Eq.~(\ref{eq:Fmax}) of the main text.

The arguments presented above can be directly extended to $n$ parties, in which case $F_{\max}$ is defined as 
\begin{equation}
F_{\max}(\rho_{n})=\max_{\Lambda\in\mathrm{GLOCC}}\mathrm{Tr}\left[\Lambda(\rho_{n})\ket{0}\!\bra{0}^{\otimes n}\right],
\end{equation}
where $\rho_n$ is an $n$-partite quantum state. Using the same arguments as above, we can express $F_{\max}$ as follows:
\begin{equation}
F_{\max}(\rho_{n})=\max_{\{U_{i}\}}\mathrm{Tr}\left[U_{1}\otimes\cdots\otimes U_{n}\rho_{n}U_{1}^{\dagger}\otimes\cdots\otimes U_{n}^{\dagger}\ket{0}\!\bra{0}^{\otimes n}\right],
\end{equation}
where the maximum is taken over all local unitaries $U_i$.

\section{Proof of Proposition~\ref{prop:optimizer-purity}}

First, we note that a pure bipartite pure state $\psi$ is close to a PPT state if and only if it is close to a separable state.
This is because
\begin{align}
\norm{\psi^{T_{B}}}_{\infty} & =\max_{\mu\in\textrm{SEP}}F(\psi,\mu)\leq\max_{\mu\in\textrm{PPT}}F(\psi,\mu)\\
 & =\max_{\mu\in\textrm{PPT}}\Tr(\psi\mu)=\max_{\mu\in\textrm{PPT}}\Tr(\psi^{T_{B}}\mu)\nonumber \\
 & \leq\max_{\norm{\mu}_{1}\leq1}\Tr(\psi^{T_{B}}\mu)=\norm{\psi^{T_{B}}}_{\infty},\nonumber 
\end{align}
where in the first step, we used the fact that taking the partial transpose in the Schmidt basis of a pure state gives us
\begin{align}
\pqty{\sum_{ij}\sqrt{p_{i}p_{j}}\ket{ii}\bra{jj}}^{T_{B}}
=& \sum_{ij} \sqrt{p_i p_j} \ket{ij} \bra{ji}
\\
=& \sum_i p_i \ketbra{ii} \nonumber
\\
&+ \sum_{i < j} \sqrt{p_i p_j} \pqty{ \frac{\ket{ij} + \ket{ji}}{\sqrt{2}} }\pqty{ \frac{\bra{ij} + \bra{ji}}{\sqrt{2}} } \nonumber
\\
&- \sum_{i < j} \sqrt{p_i p_j} \pqty{ \frac{\ket{ij} - \ket{ji}}{\sqrt{2}} }\pqty{ \frac{\bra{ij} - \bra{ji}}{\sqrt{2}} }, \nonumber
\end{align}
which means $\norm{T_B(\phi)}_{\infty}$ gives us the square of the largest Schmidt coefficient. An implication of the above results is that for any bipartite pure state $\ket{\psi}$ it holds that 
\begin{equation}
\max_{\sigma\in\mathrm{SEP}}F(\psi,\sigma)=\max_{\sigma\in\mathrm{PPT}}F(\psi,\sigma)=\lambda_{\max}(\psi^{A}). \label{eq:FPPTSEP}
\end{equation}

Consider now a PPT state $\sigma_\mathrm{PPT}$ with the property that
\begin{equation}
    \lambda_{\max}(\sigma_\mathrm{PPT}) = 1-\varepsilon. \label{eq:lambdaMax}
\end{equation}
Let $\ket{\phi}$ be the eigenvector corresponding to the largest eigenvalue of $\sigma_\mathrm{PPT}$. It holds that $F(\phi,\sigma_\mathrm{PPT})=\lambda_{\max}(\sigma_\mathrm{PPT})=1-\varepsilon$. Moreover, due to Eq.~(\ref{eq:FPPTSEP}) there is a separable state $\sigma_\mathrm{SEP}$ such that $F(\phi,\sigma_{\mathrm{SEP}})\geq1-\varepsilon$.

In the next step, consider a state $\rho$ and let $\sigma_{\mathrm{PPT}}$ be such that 
\begin{equation}
\mathrm{Tr}(\rho\sigma_{\mathrm{PPT}})\geq\max_{\sigma\in\mathrm{PPT}}\mathrm{Tr}(\rho\sigma)-\delta \label{eq:SigmaSDP}
\end{equation}
for a small $\delta > 0$. We note that an arbitrarily small $\delta > 0$ can be achieved via SDP. Moreover, we assume that $\sigma_{\mathrm{PPT}}$ fulfills Eq.~(\ref{eq:lambdaMax}) for a small $\varepsilon > 0$. Then there exists a separable state $\sigma_{\mathrm{SEP}}$ and a pure state $\ket{\phi}$ with the properties as discussed below Eq.~(\ref{eq:lambdaMax}). We further find that 
\begin{align} \label{eq:ProofPure-1}
\left|\mathrm{Tr}\left(\rho\sigma_{\mathrm{PPT}}\right)-\mathrm{Tr}\left(\rho\sigma_{\mathrm{SEP}}\right)\right| & =\left|\mathrm{Tr}\left(\rho\left[\sigma_{\mathrm{PPT}}-\sigma_{\mathrm{SEP}}\right]\right)\right|\\
 & \leq\sqrt{\mathrm{Tr}(\rho^{2})\mathrm{\mathrm{Tr}([\sigma_{\mathrm{PPT}}-\sigma_{\mathrm{SEP}}]^{2})}}\nonumber \\
 & \leq\sqrt{\mathrm{Tr}(\rho^{2})}\left\Vert \sigma_{\mathrm{PPT}}-\sigma_{\mathrm{SEP}}\right\Vert _{1}\nonumber \\
 & \leq\left\Vert \sigma_{\mathrm{PPT}}-\sigma_{\mathrm{SEP}}\right\Vert _{1}\nonumber \\
 & \leq\left\Vert \sigma_{\mathrm{PPT}}-\phi\right\Vert _{1}+\left\Vert \phi-\sigma_{\mathrm{SEP}}\right\Vert _{1}\nonumber \\
 & \leq 2 \sqrt{1-F\left(\sigma_{\mathrm{PPT}},\phi\right)}\nonumber \\
 & + 2 \sqrt{1-F\left(\phi,\sigma_{\mathrm{SEP}}\right)} \nonumber \\
 & \leq 4 \sqrt{\varepsilon}.\nonumber 
\end{align}
This result implies the following inequality:
\begin{align}
\mathrm{Tr}\left(\rho\sigma_{\mathrm{PPT}}\right) & \leq\mathrm{Tr}\left(\rho\sigma_{\mathrm{SEP}}\right)+ 4 \sqrt{\varepsilon}\\
 & \leq\max_{\sigma\in\mathrm{SEP}}\mathrm{Tr}\left(\rho\sigma\right)+ 4 \sqrt{\varepsilon}. \nonumber
\end{align}

From Eq.~(\ref{eq:SigmaSDP}) we further obtain
\begin{equation}
\mathrm{Tr}(\rho\sigma_{\mathrm{PPT}})\geq\max_{\sigma\in\mathrm{SEP}}\mathrm{Tr}(\rho\sigma)-\delta.
\end{equation}
In summary, we can bound $\max_{\sigma\in\mathrm{SEP}}\mathrm{Tr}(\rho\sigma)$ from above and from below as follows: 
\begin{equation}
\mathrm{Tr}(\rho\sigma_{\mathrm{PPT}})+\delta\geq\max_{\sigma\in\mathrm{SEP}}\mathrm{Tr}(\rho\sigma)\geq\mathrm{Tr}(\rho\sigma_{\mathrm{PPT}})- 4 \sqrt{\varepsilon}.
\end{equation}
The prof of Eq.~(\ref{eq:FmaxSDP}) of the main text is complete by recalling that $F_{\max}(\rho)=\max_{\sigma\in\mathrm{SEP}}\mathrm{Tr}(\rho\sigma)$.

We can extend the argument above to more than two parties. We will now demonstrate this for $n=3$. Here, PPT means PPT in all bipartitions, and separable means full separability.
To get a similar result as in the bipartite case, we need to bound the distance from a pure state that is almost product to a fully separable state.
Let $\psi_{ABC}$ be a tripartite pure state and $\sigma_{\mathrm{PPT}}$ is tripartite state which is PPT in all bipartitions such that $F(\psi,\sigma_{\mathrm{PPT}})=1-\varepsilon$ with some $\varepsilon > 0$. Using the same arguments as in the proof of Eq.~(\ref{eq:FmaxSDP}) of the main text, there exist product states $\phi^1_A \otimes \phi^1_{BC}$ and $\phi^2_{AB} \otimes \phi^2_C$ 
such that 
\begin{align}
F(\psi,\phi_{A}^{1}\otimes\phi_{BC}^{1}) & \geq1-\varepsilon,\\
F(\psi,\phi_{B}^{2}\otimes\phi_{AC}^{2}) & \geq1-\varepsilon.
\end{align}
Using triangle inequality for the trace norm and the Fuchs-van de Graaf inequality, we have
\begin{align}
\norm{\psi_{ABC}-\phi_{A}^{1}\otimes\phi_{B}^{2}\otimes\phi_{C}^{2}}_{1}\leq & \norm{\psi_{ABC}-\phi_{A}^{1}\otimes\phi_{BC}^{1}}_{1}\\
 & +\norm{\phi_{A}^{1}\otimes\phi_{BC}^{1}-\phi_{A}^{1}\otimes\phi_{B}^{2}\otimes\phi_{C}^{2}}_{1}\nonumber \\
= & \norm{\psi_{ABC}-\phi_{A}^{1}\otimes\phi_{BC}^{1}}_{1}\nonumber \\
 & +\norm{\phi_{BC}^{1}-\phi_{B}^{2}\otimes\phi_{C}^{2}}_{1}\nonumber \\
\leq & \norm{\psi_{ABC}-\phi_{A}^{1}\otimes\phi_{BC}^{1}}_{1}\nonumber \\
 & +\norm{\phi_{A}^{1}\otimes\phi_{BC}^{1}-\phi_{B}^{2}\otimes\phi_{AC}^{2}}_{1}\nonumber \\
\leq & \norm{\psi_{ABC}-\phi_{A}^{1}\otimes\phi_{BC}^{1}}_{1}\nonumber \\
 & +\norm{\phi_{A}^{1}\otimes\phi_{BC}^{1}-\psi_{ABC}}_{1}\nonumber \\
 & +\norm{\psi_{ABC}-\phi_{B}^{2}\otimes\phi_{AC}^{2}}_{1}\nonumber \\
\leq & 2\norm{\psi_{ABC}-\phi_{A}^{1}\otimes\phi_{BC}^{1}}_{1}\nonumber \\
 & +\norm{\psi_{ABC}-\phi_{B}^{2}\otimes\phi_{AC}^{2}}_{1}\nonumber \\
\leq & 6\sqrt{\varepsilon}.\nonumber 
\end{align}
This implies that there exists a separable state $\sigma_{\mathrm{SEP}}$ such that $||\psi-\sigma_{\mathrm{SEP}}||_{1}\leq6\sqrt{\varepsilon}$.
Using the same arguments as in Eq.~(\ref{eq:ProofPure-1}) we find 
\begin{align}
\left|\mathrm{Tr}\left(\rho\sigma_{\mathrm{PPT}}\right)-\mathrm{Tr}\left(\rho\sigma_{\mathrm{SEP}}\right)\right| & \leq\left\Vert \sigma_{\mathrm{PPT}}-\sigma_{\mathrm{SEP}}\right\Vert _{1}\\
 & \leq\left\Vert \sigma_{\mathrm{PPT}}-\psi\right\Vert _{1}+\left\Vert \psi-\sigma_{\mathrm{SEP}}\right\Vert _{1}\nonumber \\
 & \leq8\sqrt{\varepsilon}.\nonumber 
\end{align}

In summary, this means that for $n = 3$ we can bound $F_{\max}$ by maximizing $\mathrm{Tr}(\rho\sigma)$ over all states $\sigma$ which are PPT in all bipartitions. Note that this maximization is feasible via SDP. If the optimal state $\sigma_{\mathrm{PPT}}$ is such that $\lambda_{\max}(\sigma_{\mathrm{PPT}})=1-\varepsilon$, we obtain the following bound on $F_{\max}$:
\begin{equation}
\mathrm{Tr}(\rho\sigma_{\mathrm{PPT}})+\delta\geq F_{\max}(\rho)\geq\mathrm{Tr}(\rho\sigma_{\mathrm{PPT}})-8\sqrt{\varepsilon}, \label{eq:Fmax3parties}
\end{equation}
where $\delta > 0$ is the error of the SDP.

To demonstrate the efficiency of this approximation, we have sampled 6000 Haar random pure states of three qutrits and evaluated $F_{\max}(\psi^{ABC})$. Note that for any pure tripartite state $\ket{\psi}^{ABC}$ it holds that~\cite{PhysRevA.77.062317} 
\begin{equation}
F_{\max}(\psi^{ABC})=F_{\max}(\rho^{AB})=F_{\max}(\rho^{AC})=F_{\max}(\rho^{BC}),
\end{equation}
which means that we can approximate $F_{\max}$ either by approximating over bipartite or tripartite states which are PPT in all bipartitions. Evaluating the SDP, we found numerically that $\varepsilon < 1.04 \times 10^{-10}$, which means that $F_{\max}$ can be evaluated with a precision of at least $8.2\times 10^{-5}$ in these cases. In the same way, we see that the geometric entanglement of Haar random three qutrit states can be evaluated with the same precision.

\section{Proof of Theorem \ref{thm:F}}

Let us consider a slightly modified maximization problem 
\begin{equation}
F'(\rho^{AB})=\max_{Y^{AB}}\mathrm{Tr}\left[Y^{AB}\rho^{AB}\right],
\end{equation}
where the maximum is now taken over all matrices $Y^{AB}$ with the
properties 
\begin{equation}
Y^{AB}\geq0,\,\,\,\,\,Y^{T_{B}}\geq0,\,\,\,\,\,Y^{A} = \frac{1}{d_{A}}\openone_{A}.\label{eq:AssistedSDP-4}
\end{equation}
It is clear that the maximum on the right-hand side of Eq.~(\ref{eq:AssistedSDP-3}) of the main text
can be expressed as 
\begin{equation}
\max_{X^{AB}}\mathrm{Tr}[X^{AB}\rho^{AB}]=d_{A}F'(\rho^{AB}).
\end{equation}
Moreover, $Y$ can be considered as a density matrix since $\mathrm{Tr}(Y)=1$. For $d_{A}d_{B}\leq6$
Eq. (\ref{eq:AssistedSDP-4}) implies that $Y$ is separable~\cite{HORODECKI19961},
i.e., can be written as $Y=\sum_{i}p_{i}Y_{i}^{A}\otimes Y_{i}^{B}$
with local density matrices $Y_{i}^{A,B}$ such that $\sum_{i}p_{i}Y_{i}^{A}=\openone_{A}/d_{A}$.
We thus obtain 
\begin{equation}
F'(\rho^{AB})=\max\sum_{i}\mathrm{Tr}\left[p_{i}Y_{i}^{A}\otimes Y_{i}^{B}\rho^{AB}\right],
\end{equation}
where the maximum is now taken over all local density matrices $Y_{i}^{A,B}$
and all probability distributions $p_{i}$ with $\sum_{i}p_{i}Y_{i}^{A}=\openone_{A}/d_{A}$.
Comparing this expression to Eq.~(\ref{eq:Pa-1}) of the main text, we see that $F_{\rightarrow}^{A|B}(\rho^{AB})=d_{A}F'(\rho^{AB})$.
This completes the proof.

\section{Proof of Proposition~\ref{prop:assisted-sdp}}
We will now show that $F^{A|B}$ can be bounded below by Eq.~(\ref{eq:AssistedSDP-5}) of the main text, where the maximization is done over all matrices $X^{AB}$ as defined in Eq.~(\ref{eq:AssistedSDP-6}). For this, we note that $F^{A|B}$ can be upper bounded as follows:
\begin{align}
F^{A|B}(\rho^{AB}) & =\max_{\Lambda\in\text{GLOCC}}\mathrm{Tr}\left[\Lambda\left(\rho^{AB}\right)\openone_{A}\otimes\ket{0}\!\bra{0}^{B}\right]\\
 & \leq\max_{\Lambda\in\text{GSEP}}\mathrm{Tr}\left[\rho^{AB}\Lambda\left(\openone_{A}\otimes\ket{0}\!\bra{0}^{B}\right)\right],\nonumber 
\end{align}
where GSEP denotes the set of separable operations which preserve the maximally mixed state. For any $\Lambda \in \mathrm{GSEP}$ it holds that 
\begin{align}
\openone_{AB}\geq\Lambda\left(\openone_{A}\otimes\ket{0}\!\bra{0}^{B}\right) & \geq0,\\
\mathrm{Tr}\left[\Lambda\left(\openone_{A}\otimes\ket{0}\!\bra{0}^{B}\right)\right] & =2,
\end{align}
and moreover $\Lambda(\openone_{A}\otimes\ket{0}\!\bra{0}^{B})$ has positive partial transpose. This completes the proof.

\section{Proof of Theorem \ref{thm:Complementarity}}

Let $\{M_{i}^{A}\}$ be an optimal POVM on Alice's side, maximizing
Bob's purity. Recall that the probability $p_{i}$ of the outcome
$i$ as well as the post-measurement state on Bob's side are given
by 
\begin{align}
p_{i} & =\mathrm{Tr}[M_{i}^{A}\otimes\openone^{B}\rho^{AB}],\\
\sigma_{i}^{B} & =\frac{1}{p_{i}}\mathrm{Tr}_{A}[M_{i}^{A}\otimes\openone^{B}\rho^{AB}].
\end{align}
Denoting by $\ket{\nu_{i}}^{B}$ the eigenstate corresponding to the
largest eigenvalue of $\sigma_{i}^{B}$ we can express $F_{\rightarrow}^{A|B}$
as follows: 
\begin{equation}
F_{\rightarrow}^{A|B}(\rho^{AB})=\sum_{i}\mathrm{Tr}\left[M_{i}^{A}\otimes\nu_{i}^{B}\rho^{AB}\right]=\sum_{i}p_{i}\lambda_{\max}(\sigma_{i}^{B}),
\end{equation}
where $\lambda_{\max}(\rho)$ denotes the largest eigenvalue of $\rho$.
Without loss of generality we can further assume that each POVM element
$M_{i}^{A}$ has rank one.

Let us now consider the action of the POVM $\{M_{i}^{A}\}$ on the
state $\ket{\psi}^{ABC}$, which is a purification of $\rho^{AB}$.
Conditioned on the measurement outcome $i$, Bob and Charlie end up
with the state 
\begin{equation}
\sigma_{i}^{BC}=\frac{1}{p_{i}}\mathrm{Tr}_{A}\left[M_{i}^{A}\otimes\openone^{BC}\psi^{ABC}\right].
\end{equation}
Since $M_{i}^{A}$ have rank one, each state $\sigma_{i}^{BC}$ is
a pure extension of the state $\sigma_{i}^{B}$. Recall that the geometric
entanglement of a pure state $\sigma_{i}^{BC}$ is related to the
largest eigenvalue of the reduced state $\sigma_{i}^{B}$ as follows~\cite{Shimony1995,PhysRevA.68.042307}: 
\begin{equation}
E_{g}(\sigma_{i}^{BC})=1-\lambda_{\max}(\sigma_{i}^{B}).
\end{equation}
Combining these results, we thus obtain: 
\begin{equation}
F_{\rightarrow}^{A|B}(\rho^{AB})=1-\sum_{i}p_{i}E_{g}(\sigma_{i}^{BC}).
\end{equation}
Using convexity of the geometric entanglement~\cite{PhysRevA.68.042307}
we further arrive at 
\begin{equation}
F_{\rightarrow}^{A|B}(\rho^{AB})\leq1-E_{g}(\rho^{BC}).
\end{equation}

To complete the proof of the theorem it remains to show that this
inequality is actually an equality. To see this, note that there always
exist a pure state decomposition $\{q_{i},\ket{\phi_{i}}^{BC}\}$
of the state $\rho^{BC}$ such that 
\begin{equation}
E_{g}(\rho^{BC})=\sum_{i}q_{i}E_{g}(\phi_{i}^{BC})=1-\sum_{i}q_{i}\lambda_{\max}(\phi_{i}^{B}).
\end{equation}
Moreover, there always exists a POVM $\{M_{i}^{A}\}$ on Alice's side,
such that Bob and Charlie end up with this optimal decomposition,
i.e., their post-measurement state corresponds to $\phi_{i}^{BC}$
with probability $q_{i}$~\cite{schrodinger_1936,HUGHSTON199314}. If Alice applies this POVM on
her part of $\rho^{AB}$, then the post-measurement state of Bob is
given by $\phi_{i}^{B}$ with probability $q_{i}$. Correspondingly,
the purity which Bob can obtain is given by $\sum_{i}q_{i}\lambda_{\max}(\phi_{i}^{B})$,
which is equivalent to $1-E_{g}(\rho^{BC})$. This completes the proof. 

\section{SDP lower bound on geometric entanglement}

\begin{figure}
\begin{centering}
\includegraphics[width=\columnwidth]{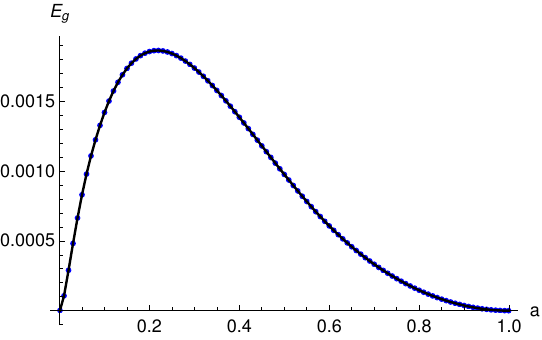}
\par\end{centering}
\caption{\label{fig:BE3} Bounds on geometric entanglement for the $2 \times 4$ PPT entangled states $\rho_a$. Solid curve shows the lower bound on $E_g$ from Eq.~(\ref{eq:EgLowerBound-1}), dots represent an upper bound obtained via the algorithm described in~\cite{PhysRevA.84.022323}.}
\end{figure}

\begin{figure}
\begin{centering}
\includegraphics[width=\columnwidth]{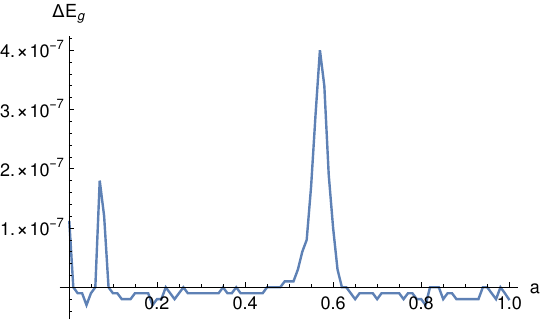}
\par\end{centering}
\caption{\label{fig:BE2} Numerical gap between the upper and the lower bound on the geometric entanglement for the states $\rho_a$.}
\end{figure}

\begin{figure*}
\begin{centering}
\includegraphics[width=0.9\columnwidth]{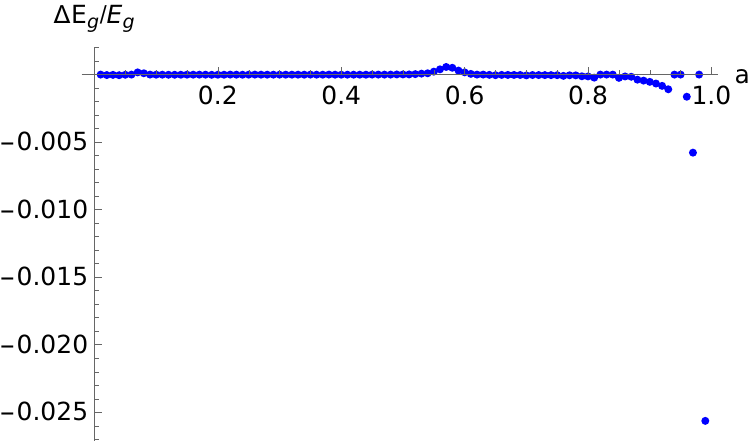} \hspace{1cm} \includegraphics[width=0.9\columnwidth]{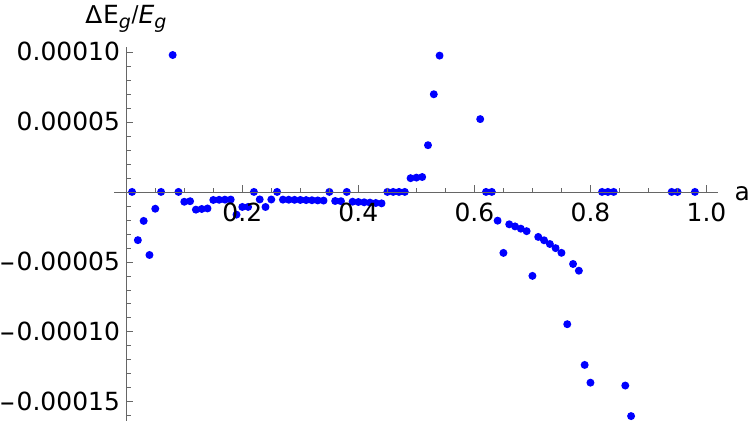}
\par\end{centering}
\caption{\label{fig:BE4} Relative numerical gap $\Delta E_g/E_g$ for the states $\rho_a$ in the ranges $-0.025 \leq \Delta E_g/E_g \leq 0.002$ (left part of the figure) and $-0.00015 \leq \Delta E_g/E_g \leq 0.0001$ (right part of the figure).}
\end{figure*}

We will now demonstrate how the results presented in our article can be used for bounding the geometric entanglement for bipartite quantum states. As stated in the main text, from Theorem~\ref{thm:Complementarity} we get the following lower bound on the geometric entanglement
\begin{align}
E_{g}(\rho^{BC}) & \geq1-\max_{X^{AB}}\mathrm{Tr}\left[X^{AB}\rho^{AB}\right], \label{eq:EgLowerBound}
\end{align}
where $\ket{\psi}^{ABC}$ is a purification of both states $\rho^{AB}$ and $\rho^{BC}$, and the maximum is taken over all matrices $X^{AB}$ with the properties 
\begin{equation}
X^{AB}\geq0,\quad X^{T_{B}}\geq0,\quad X^{A} = \openone_{A}.
\end{equation}

By similar arguments, it is possible to show that $E_g$ can also be bounded from below as follows:
\begin{equation}
E_{g}(\rho^{BC})\geq1-\max_{X^{ABC}}\mathrm{Tr}\left[X^{ABC}\psi^{ABC}\right],\label{eq:EgLowerBound-1}
\end{equation}
where $\ket{\psi}^{ABC}$ is a purification of $\rho^{BC}$ and the maximum is taken over all $X^{ABC}$ with the properties
\begin{equation}
X^{ABC}\geq0,\,\,\,X^{T_{A}}\geq0,\,\,\,X^{T_{B}}\geq0,\,\,\,X^{T_{C}}\geq0,\,\,\,X^{A}=\openone_{A}. \label{eq:EgLowerBoundImproved}
\end{equation}
To prove Eq.~(\ref{eq:EgLowerBound-1}), note that the geometric entanglement can be expressed as follows~\cite{Streltsov_2010}:
\begin{equation}
E_{g}(\rho^{BC})=1-\max\sum_{i}p_{i}F_{s}(\psi_{i}^{BC}), \label{eq:EgFs}
\end{equation}
where the maximum is taken over all pure state decompositions of $\rho^{BC}$, and $F_{s}(\rho)=\max_{\sigma\in\mathrm{SEP}}F(\rho,\sigma)$. From Eq.~(\ref{eq:EgFs}) it follows that there exist some POVM $\{M_{i}^{A}\}$ and pure local states $\ket{\mu_i}^B$ and $\ket{\nu_i}^C$ such that 
\begin{equation}
E_{g}(\rho^{BC})=1-\mathrm{Tr}\left[\sum_{i}M_{i}^{A}\otimes\ket{\mu_{i}}\!\bra{\mu_{i}}^{B}\otimes\ket{\nu_{i}}\!\bra{\nu_{i}}^{C}\ket{\psi}\!\bra{\psi}^{ABC}\right].
\end{equation}
Since the set of matrices $X^{ABC}$ with the properties~(\ref{eq:EgLowerBoundImproved}) contains all matrices of the form $\sum_{i}M_{i}^{A}\otimes\ket{\mu_{i}}\!\bra{\mu_{i}}^{B}\otimes\ket{\nu_{i}}\!\bra{\nu_{i}}^{C}$, we arrive at Eq.~(\ref{eq:EgLowerBound-1}). We note that in the same way it is also possible to construct lower bounds on the multipartite geometric entanglement.

An important family of states for which the evaluation of entanglement measures is notoriously difficult are PPT entangled states. Important examples in this context are the $2 \times 4$ states~\cite{HORODECKI1997333}
\begin{equation}
\rho_{a}=\frac{1}{7a+1}\left(\begin{array}{cccccccc}
a & 0 & 0 & 0 & 0 & a & 0 & 0\\
0 & a & 0 & 0 & 0 & 0 & a & 0\\
0 & 0 & a & 0 & 0 & 0 & 0 & a\\
0 & 0 & 0 & a & 0 & 0 & 0 & 0\\
0 & 0 & 0 & 0 & \frac{1+a}{2} & 0 & 0 & \frac{\sqrt{1-a^{2}}}{2}\\
a & 0 & 0 & 0 & 0 & a & 0 & 0\\
0 & a & 0 & 0 & 0 & 0 & a & 0\\
0 & 0 & a & 0 & \frac{\sqrt{1-a^{2}}}{2} & 0 & 0 & \frac{1+a}{2}
\end{array}\right),
\end{equation}
which are PPT entangled in the range $0 < a < 1$. Solid curve in Fig.~\ref{fig:BE3} shows the lower bound from Eq.~(\ref{eq:EgLowerBound-1}). An upper bound on the geometric entanglement of mixed states can be obtained using the methods described in~\cite{PhysRevA.84.022323}. We have evaluated the upper bound for $a=n/100$, with the integer $n$ ranging from 0 to 100. The result is shown as dots in Fig.~\ref{fig:BE3}. The gap between the upper and the lower bound is shown in Fig.~\ref{fig:BE2}. The numerical gap $\Delta E_g$ is never larger than $4\times10^{-7}$, and the average gap is approximately $\overline{\Delta E_{g}}\approx1.2\times10^{-8}$. In Fig.~\ref{fig:BE4} we also show the relative numerical gap $\Delta E_g/E_g$. In the vicinity of separable states, i.e., for $a=0.001$ and $a=0.999$ we further obtained $\Delta E_g/E_g = 0.06$ and $\Delta E_g/E_g = -2$, respectively. These values are not shown in Fig.~\ref{fig:BE4}.

\section{Evaluating geometric entanglement for multipartite pure states}
We note that the evaluation of $F_{\max}$ can also be used to evaluate the geometric entanglement for multipartite pure states. We will now discuss this explicitly for 3 and 4 parties. These results follow directly from~\cite{PhysRevA.77.062317}, and we present them here for completeness.

In more detail, for a 3-partite pure state $\ket{\psi}^{ABC}$ it holds that 
\begin{align}
E_{g}(\ket{\psi}^{ABC}) & =1-F_{\max}(\rho^{AB})=1-F_{\max}(\rho^{BC})\\
 & =1-F_{\max}(\rho^{AC})=1-F_{\max}(\psi^{ABC}),\nonumber 
\end{align}
where $\rho^{XY}$ denote the reduced states of $\ket{\psi}$. This directly implies that the SDP for evaluating $F_{\max}$ in bipartite states can be used to evaluate $E_g$ for tripartite pure states. For $d_Ad_B \leq 6$ it is possible do evaluate $F_{\max}$ via SDP, which means that $E_g$ is also feasible via SDP in all such cases. 

For more general sets of 3-partite pure states it is possible to estimate $E_g$ via estimation of $F_{\max}$ of the bipartite subsystems with the precision given in Eq.~(\ref{eq:FmaxSDP}) of the main text. Similarly, geometric entanglement for 4-partite pure states can be estimated by estimating $F_{\max}$ of the reduced 3-partite states, with precision given in Eq.~(\ref{eq:Fmax3parties}).

\end{document}